\documentclass[12pt,3p,sort&compress]{elsarticle}

\usepackage{graphicx} 
\usepackage{multirow}
\usepackage{array} 
\usepackage{amsmath}
\usepackage{amssymb}

\usepackage[colorlinks=false,linktocpage=true]{hyperref}
\usepackage[usenames]{color}
\usepackage[T1]{fontenc}

\newcommand{\pit}{\tilde{\pi}}

\def\lesssim{\mathrel{\rlap{\lower4pt\hbox{$\sim$}}
    \raise1pt\hbox{$<$}}} % less than or approx. symbol
\def\gtrsim{\mathrel{\rlap{\lower4pt\hbox{$\sim$}}
    \raise1pt\hbox{$>$}}} % greater than or approx. symbole
    
\begin{document}

\begin{frontmatter}

\title{Anisotropic Hydrodynamics:  Motivation and Methodology}

\author[kent]{Michael Strickland} 
\address[kent]{Physics Department, Kent State University, OH 44242 United States}

\begin{abstract}
In this proceedings contribution I review recent progress in our understanding of the bulk dynamics of 
relativistic systems that possess potentially large local rest frame momentum-space anisotropies.
In order to deal with these momentum-space anisotropies, a reorganization of relativistic viscous hydrodynamics
can been made around an anisotropic background, and the resulting dynamical framework has been dubbed ``anisotropic hydrodynamics.''
I also discuss expectations for the degree of momentum-space anisotropy of the quark gluon plasma
generated in relativistic heavy ion collisions at RHIC and LHC from second-order viscous hydrodynamics,
strong-coupling approaches, and weak-coupling approaches.
\end{abstract}


%\pacs{11.15.Bt, 11.10.Wx, 12.38.Mh, 25.75.-q, 52.27.Ny, 52.35.-g}

\end{frontmatter}

%%%%%%%%%%%%%%%%%%%%%%%%%%%%%%%%%%%%%%%%%%%%%%%%%%%%%%%%%%%%%%%%%%%%%%%%%%%%%%%%%%%%%%%%%%%%%%%%%%%%%%%%%%%%%%
\section{Introduction}
%%%%%%%%%%%%%%%%%%%%%%%%%%%%%%%%%%%%%%%%%%%%%%%%%%%%%%%%%%%%%%%%%%%%%%%%%%%%%%%%%%%%%%%%%%%%%%%%%%%%%%%%%%%%%%

Much has been learned, both on the theoretical and phenomenological fronts since the first $\sqrt{s_{\rm NN}} = 200$ MeV Au-Au data were made available from Au-Au collisions at the Relativistic Heavy Ion Collider (RHIC) at Brookhaven National Lab over a decade ago.  Since then the heavy ion community has collected a tremendous amount of experimental data and our theoretical understanding, both in terms of our ability to simulate the non-abelian dynamics of the quark gluon plasma (QGP) from first principles and to model the QGP based on effective models, has advanced tremendously.  Additionally, with the turn on of the Large Hadron Collider (LHC) at the European Center for Nuclear Research (CERN) in 2008, we now have access to $\sqrt{s_{\rm NN}} = 2.76$ TeV Pb-Pb data which allows us to further push into the QGP part of the phase diagram of quantum chromodynamics (QCD).  In the future, the full energy Pb-Pb runs with $\sqrt{s_{\rm NN}} = 5.5$ TeV will push us even further into the QGP phase.  In recent years a somewhat surprising twist in the QGP story has occurred that implies that the ``most perfect fluid ever generated'' possesses potentially large momentum-space anisotropies in the local rest frame (LRF).  As a result, the pressures transverse to and longitudinal to the beam line can be quite different, particularly at early times and/or near the transverse/longitudinal edges of the plasma.  This has important implications for both the dynamics and signatures of the QGP.

Information about the degree of isotropy of the QGP generated in relativistic heavy ion collisions came first from phenomenological fits to RHIC data using ideal hydrodynamics.  Since ideal hydrodynamics assumes a priori that the QGP is completely isotropic, the heavy-ion community interpreted the ability of such models to describe the $p_T$-dependence of the transverse elliptical flow as solid evidence that the QGP created in heavy ion collisions became isotropic and thermal at approximately 0.5 - 1 fm/c after the initial nuclear impact \cite{Huovinen:2001cy,Hirano:2002ds,Kolb:2003dz}.  Stepping forward, we now understand that at least the conclusion concerning isotropy must be relaxed based on results from modern viscous hydrodynamics simulations.  Since the early days of ideal hydrodynamics there has been a concerted effort to make hydrodynamical models more realistic by including the effect of shear and bulk viscosities (relaxation times).  This has lead to a proper formulation of relativistic viscous hydrodynamics \cite{Muronga:2003ta,Baier:2006um,Baier:2007ix,Dusling:2007gi,Luzum:2008cw,Song:2008hj,Denicol:2010xn,Schenke:2010rr,Schenke:2011tv,Shen:2011eg,Bozek:2011wa,Niemi:2012ry,Bozek:2012qs,Denicol:2012cn,PeraltaRamos:2012xk} and, recently, anisotropic relativistic viscous hydrodynamics \cite{Martinez:2010sc,Florkowski:2010cf,Ryblewski:2010bs,Martinez:2010sd,Ryblewski:2011aq,Florkowski:2011jg,Martinez:2012tu,Ryblewski:2012rr,Florkowski:2012as,PeraltaRamos:2012xk,Florkowski:2013uqa,Bazow:2013ifa}.  The conclusion one reaches from dissipative hydrodynamics approaches is that the QGP created in ultrarelativistic heavy ion collisions (URHICs) has quite different longitudinal (along the beam line) and transverse pressures, particularly at times $\tau \lesssim 2$ fm/c.  The strength of these anisotropies increases as one goes to early times or lower (local) initial temperatures.

In addition to the progress made in dissipative hydrodynamical modeling of the QGP, there have been significant advances in our understanding of the underlying quantum field theory processes driving the thermalization and (an-)isotropization of the QGP in the weak and strong coupling limits (see Ref.~\cite{Strickland:2013uga} for a recent review).  The picture emerging from these advances seems to fit nicely into the picture emerging from the aforementioned dissipative hydrodynamics findings, namely that the QGP created in URHICs possesses large momentum-space anisotropies in the LRF.  On the separate issue of thermalization, there is evidence from simulations of weak-coupling non-abelian dynamics that one can achieve rapid apparent longitudinal thermalization of the QGP due to the chromo-Weibel instability~\cite{Rebhan:2008uj,Attems:2012js} (see also the early time spectra reported in Ref.~\cite{Fukushima:2011nq}).  On the strong coupling front, practitioners are now able to use numerical GR to describe the formation of an extra-dimensional black hole (or more accurately an apparent horizon), which is the criterium for QGP thermalization in the AdS/CFT framework.  In an expanding background corresponding to the (approximately) boost-invariant Bjorken-like expansion of the QGP, these studies find thermalization times that are less than 1 fm/c, however, the state which emerges is momentum-space anisotropic even in the infinite 't Hooft coupling limit.

%%%%%%%%%%%%%%%%%%%%%%%%%%%%%%%%%%%%%%%%%%%%%%%%%%%%%%%%%%%%%%%%%%%%%%%%%%%%%%%%%%%%%%%%%%%%%%%%%%%%%%
\section{Momentum-space Anisotropies in the QGP}
%%%%%%%%%%%%%%%%%%%%%%%%%%%%%%%%%%%%%%%%%%%%%%%%%%%%%%%%%%%%%%%%%%%%%%%%%%%%%%%%%%%%%%%%%%%%%%%%%%%%%%

As discussed above, many disparate approaches to QGP dynamics consistently find that the QGP created in URHICs possesses LRF momentum-space anisotropies in the $p_T$-$p_L$ plane.  As the first indication of this, let's consider relativistic viscous hydrodynamics for a system that is transversely homogenous and boost invariant in the longitudinal direction, aka 0+1d dynamics.  In this case, first-order Navier Stokes (NS) viscous hydrodynamics predicts that the LRF shear correction to the ideal pressures is diagonal with space-like components $\pi^{zz} = - 4\eta/3\tau = -2\pi^{xx} = - 2\pi^{yy}$, where $\eta$ is the shear viscosity and $\tau$ is the proper time.  In viscous hydrodynamics, the longitudinal pressure is given by ${\cal P}_L = P_{\rm eq} + \pi^{zz}$ and the transverse pressure by ${\cal P}_T =  P_{\rm eq} + \pi^{xx}$.  Assuming an ideal equation of state (EoS), the resulting ratio of the longitudinal pressure over the transverse pressure from first order viscous hydrodynamics can be expressed as
\begin{equation}
\left(\frac{{\cal P}_L}{{\cal P}_T}\right)_{\rm NS} = \frac{ 3 \tau T - 16\bar\eta }{ 3 \tau T + 8\bar\eta } \, ,
\label{eq:aniso}
\end{equation}
where $\bar\eta \equiv \eta/{\cal S}$ with ${\cal S}$ being the entropy density.  Assuming RHIC-like initial conditions with $T_0 = 400$ MeV at $\tau_0 = 0.5$ fm/c and taking the conjectured lower bound $\bar\eta = 1/4\pi$ \cite{Policastro:2001yc}, one finds $\left({\cal P}_L/{\cal P}_T\right)_{\rm NS} \simeq 0.5$.  For LHC-like initial conditions with $T_0 = 600$ MeV at $\tau_0 = 0.25$ fm/c and once again taking $\bar\eta = 1/4\pi$ one finds $\left({\cal P}_L/{\cal P}_T\right)_{\rm NS} \simeq 0.35$.  This means that even in the best case scenario of $\bar\eta = 1/4\pi$, viscous hydrodynamics itself predicts rather sizable momentum-space anisotropies.  For larger values of $\bar\eta$, one obtains even larger momentum-space anisotropies.  In addition, one can see from Eq.~(\ref{eq:aniso}) that, at fixed initial proper time, the level of momentum-space anisotropy increases as one lowers the temperature.  This means, in practice, that as one moves away from the center of the nuclear overlap region towards the transverse edge, the level of momentum-space anisotropy increases.

\begin{figure}[t]
\begin{center}
\includegraphics[width=0.45\textwidth]{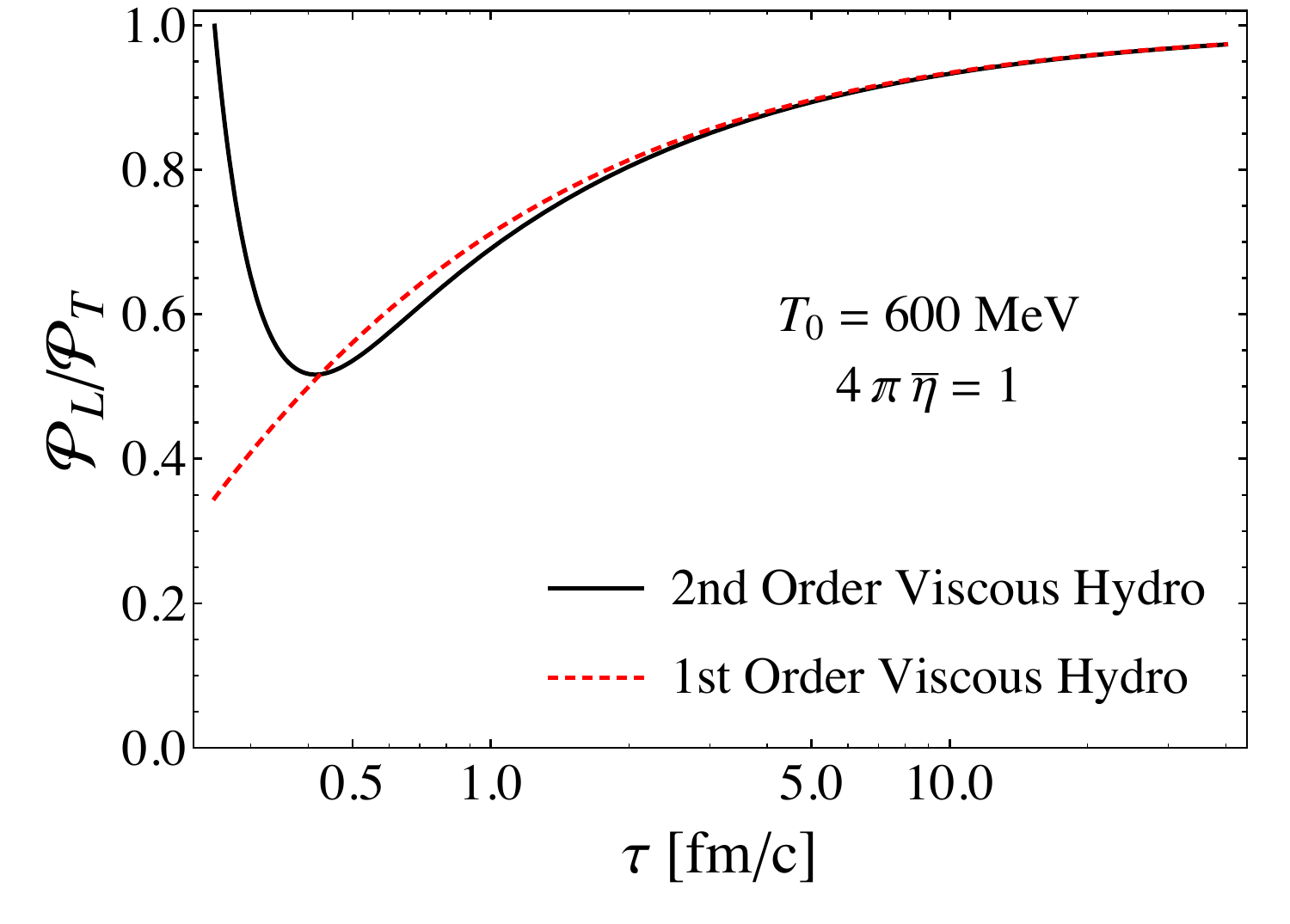}
\includegraphics[width=0.45\textwidth]{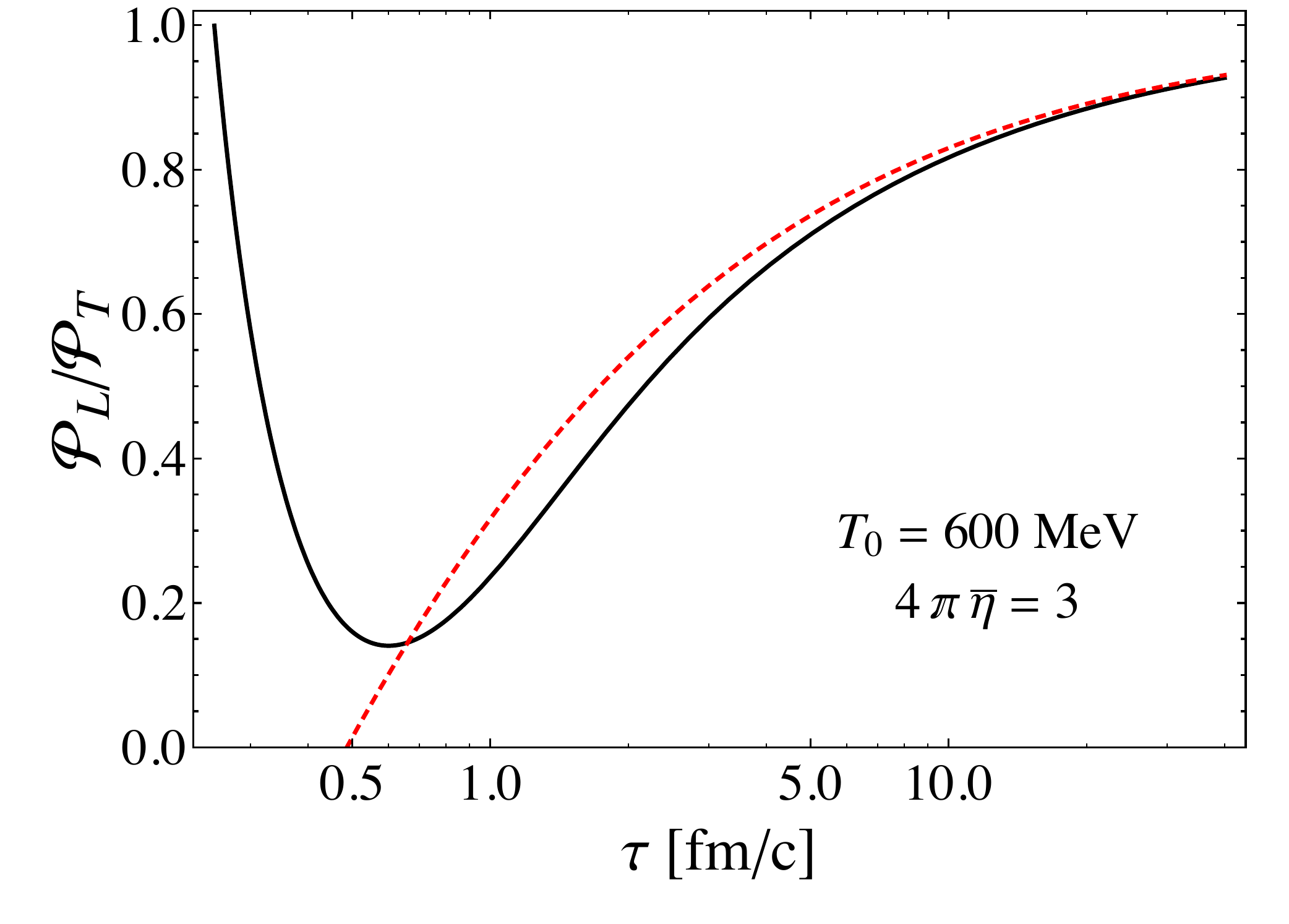}\\
\includegraphics[width=0.45\textwidth]{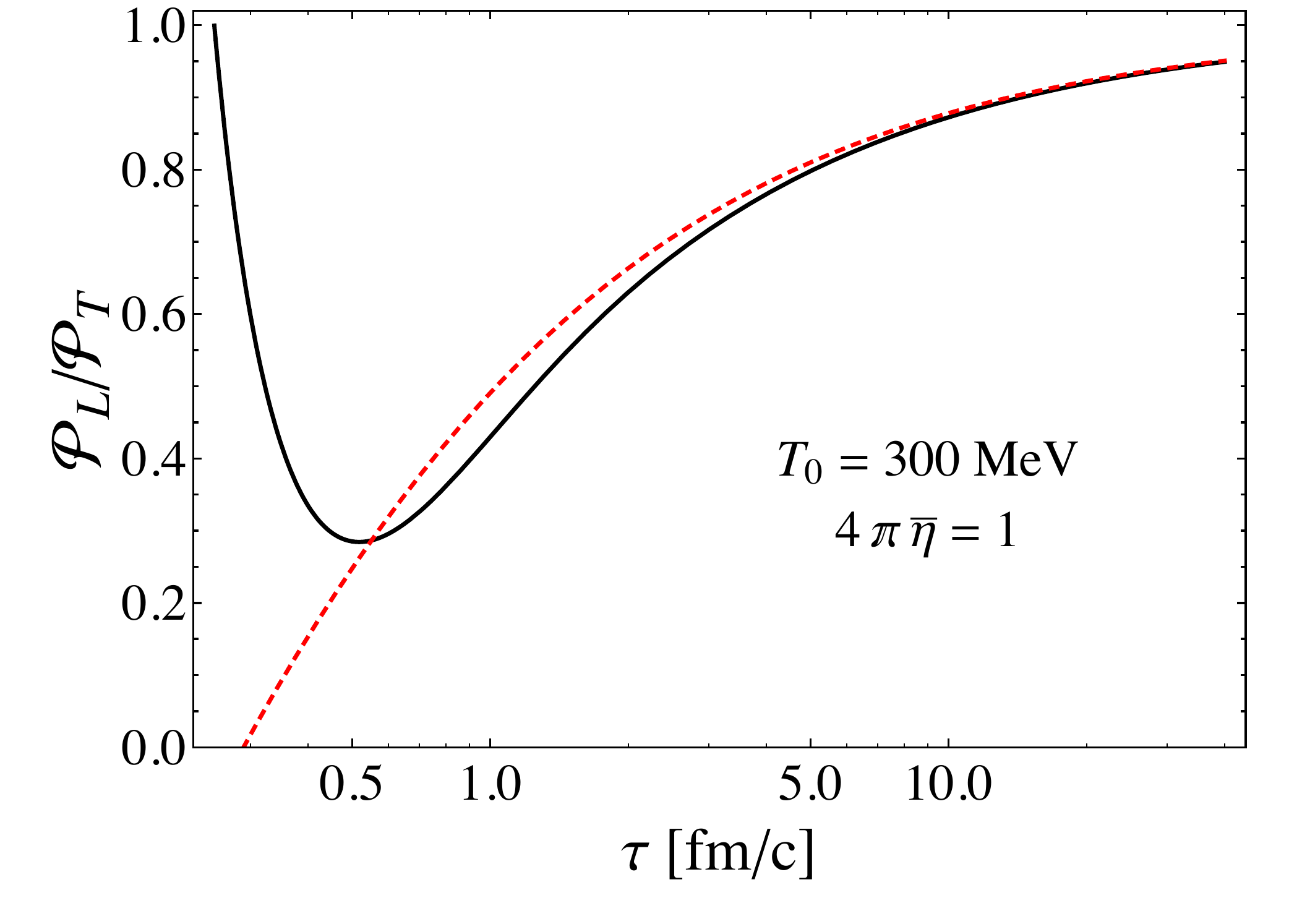}
\includegraphics[width=0.45\textwidth]{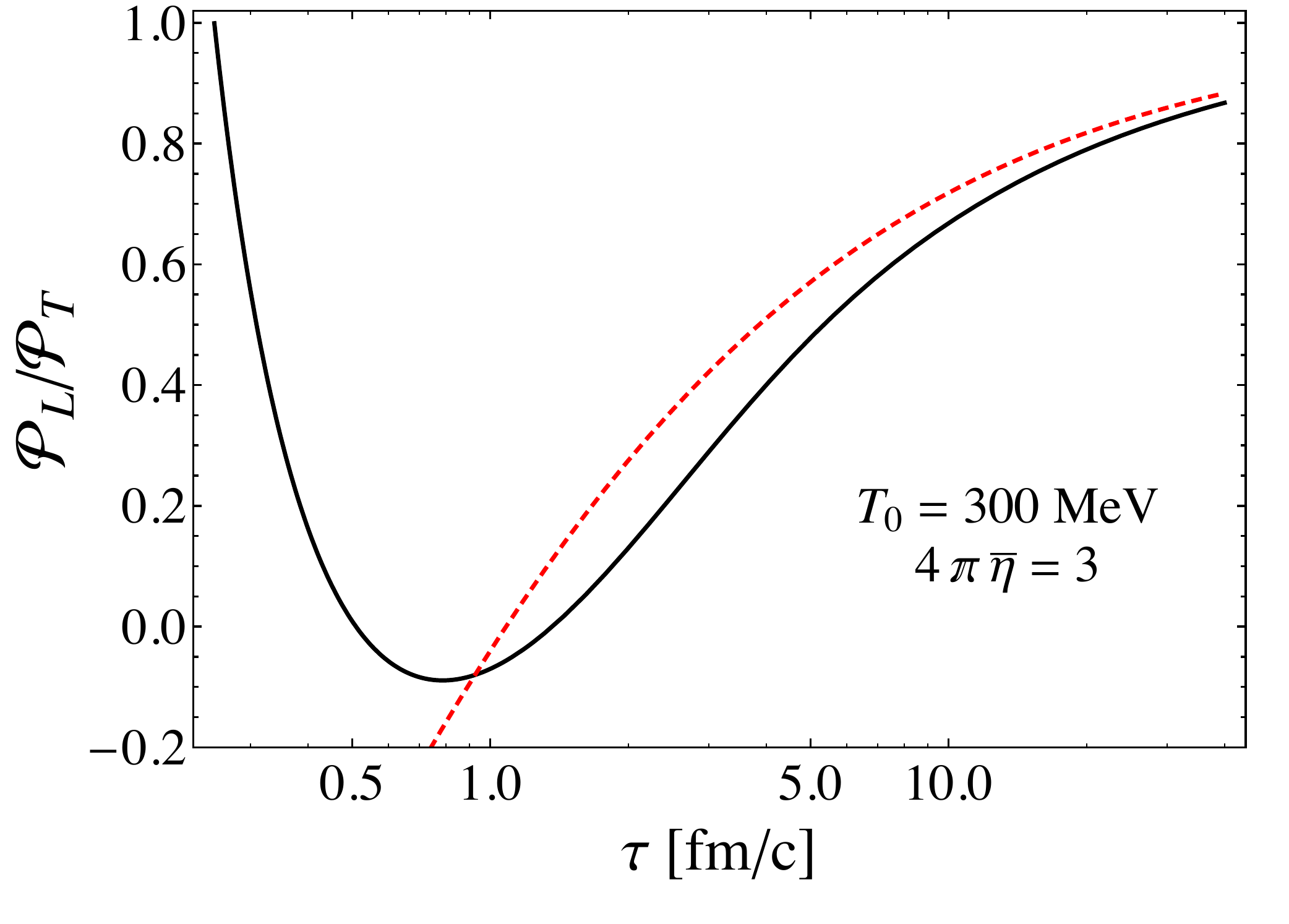}
\end{center}
\vspace{-7mm}
\caption{Pressure anisotropy as a function of proper time assuming an initially isotropic system with $T_0 = 600$ MeV (top row) and $T_0 = 300$ MeV (bottom row) at $\tau_0 =$ 0.25 fm/c for $4\pi\bar\eta =$ 1 (left column) and 3 (right column).  Solid black line is the solution of the second order coupled differential equations and the red dashed line is the first-order ``Navier-Stokes'' solution.}
\label{fig:nscomp}
\end{figure}

Of course, since first-order relativistic viscous hydrodynamics is acausal, the analysis above is not the full story.  It does, however, provide important intuitive guidance since the causal second-order version of the theory has the first-order solution as an attractive ``fixed point'' of the dynamics.  Because of this, one expects large momentum-space anisotropies to emerge within a few multiples of the shear relaxation time $\tau_\pi$.  In the strong coupling limit of ${\cal N}=4$ SYM one finds $\tau_\pi = (2 - \log 2)/2 \pi T$ \cite{Baier:2007ix,Bhattacharyya:2008jc} which gives $\tau_\pi \sim 0.1$ fm/c and $\tau_\pi \sim 0.07$ fm/c for the RHIC- and LHC-like initial conditions stated above, respectively.\footnote{A similar time scale emerges within the kinetic theory framework.}   To demonstrate this quantitatively, in Fig.~\ref{fig:nscomp} I plot the solution of the second order Israel-Stewart 0+1d viscous hydrodynamical equations assuming an isotropic initial condition and the NS solution together.  In the left column I assumed $4\pi\bar\eta = 1$ and in the right column I assumed $4\pi\bar\eta = 3$ ($\bar\eta \simeq 0.24$) with $\tau_\pi = 2 (2 - \log 2) \bar\eta/T$ in both cases.  As can be seen from this figure, even if one starts with an isotropic initial condition, within a few multiples of the shear relaxation time one approaches the NS solution, overshoots it, and then approaches it from below.  The value of $\bar\eta$ in the right column is approximately the same as that extracted from recent fits to LHC collective flow data.  I note that if one further increases $\bar\eta$, then one finds negative longitudinal pressures in second-order viscous hydrodynamics as well.  This can be seen in the lower right panel of Fig.~\ref{fig:nscomp}.

Based on the preceding discussion one learns the value of $\bar\eta$ extracted from LHC data \cite{Gale:2012rq} implies that the system may be highly momentum-space anisotropic with the momentum-space anisotropies persisting throughout the evolution of the QGP.  However, before drawing conclusions based solely on the relativistic viscous hydrodynamics, we can ask the corresponding question within the context of the AdS/CFT framework.  Several groups have been working on methods to address the question of early-time dynamics within the context of the AdS/CFT framework.  Here I focus on the work of two groups:  Heller et al. \cite{Heller:2011ju} and van der Schee et al. \cite{vanderSchee:2013pia} who both simulated the dynamics of an expanding QGP using numerical general relativity (GR).  In the work of Heller et al. they simulated the early time dynamics of a 0+1d system by numerically solving the GR equations in the bulk.  In the work of van der Schee et al. \cite{vanderSchee:2013pia} they performed similar numerical GR evolution but in the case of a 1+1d radially symmetric system including transverse expansion.  Both of these studies found early-time pressure anisotropies on the order of ${\cal P}_L/{\cal P}_T \sim 0.31$ or smaller.   Since these results were obtained in the context of the strong-coupling limit which implies $4\pi\bar\eta=1$, the pressure anisotropy found is an upper bound on what to expect in reality.

Having covered the degree of momentum-space anisotropy predicted by viscous hydrodynamics and the AdS/CFT approach, I would now like to briefly discuss the pressure anisotropies expected within the Color Glass Condensate (CGC) \cite{McLerran:1993ni,Iancu:2003xm} framework and weakly-coupled gauge field theory in general.  In the CGC framework, the fields are boost-invariant to first approximation.  As a result, the leading order prediction is that longitudinal pressure is zero.\footnote{At $\tau = 0^+$, the longitudinal pressure is negative due to coherent field effects; however, within a few fractions of a fm/c it becomes positive and at leading order goes to zero rapidly.}  Including finite energy corrections results in a very small longitudinal pressure.  Currently, it is believed that the primary driving force for restoring isotropy in the gauge field sector are plasma instabilities such as the chromo-Weibel instability \cite{Strickland:2007fm}; however, so far practitioners have found that, even taking into account the unstable gauge field dynamics, the timescale for isotropization in classical Yang-Mills simulations is very long \cite{Romatschke:2006nk,Berges:2007re}.  The recent work of Epelbaum and Gelis \cite{Gelis:2013rba} has included resummation of next-to-leading order (NLO) quantum loop corrections to initial CGC fluctuations, and simulations in this framework find early-time pressure anisotropies on the order of $0.01\,$-$\,0.5$, depending on the assumed value of the strong coupling constant $g_s = 0.1\,$-$\,0.4$.  In the context of hard-loop simulations of chromo-Weibel instability evolution, one finds rapid thermalization of the plasma in the sense that a Boltzmann distribution of gluon modes is established within $\sim$ 1 fm/c; however, large pressure anisotropies persist for at least $5\,$-$\,6$ fm/c \cite{Rebhan:2008uj,Attems:2012js}.

%%%%%%%%%%%%%%%%%%%%%%%%%%%%%%%%%%%%%%%%%%%%%%%%%%%%%%%%%%%%%%%%%%%%%%%%%%%%%%%%%%%%%%%%%%%%%%%%%%%%%%
\section{Anisotropic Hydrodynamics}
%%%%%%%%%%%%%%%%%%%%%%%%%%%%%%%%%%%%%%%%%%%%%%%%%%%%%%%%%%%%%%%%%%%%%%%%%%%%%%%%%%%%%%%%%%%%%%%%%%%%%%

As pointed out above, the assumption that the system is nearly isotropic in the LRF breaks down at early times and near the ``edges'' of the system.
To account for these large early-time deviations from local momentum isotropy non-perturbatively, a framework called ``anisotropic hydrodynamics'' was developed \cite{Martinez:2010sc,Florkowski:2010cf,Ryblewski:2010bs,Martinez:2010sd,Ryblewski:2011aq,Florkowski:2011jg,Martinez:2012tu,Ryblewski:2012rr,Florkowski:2012as,Florkowski:2013uqa}. Anisotropic hydrodynamics extends traditional viscous hydrodynamical treatments to cases in which the local transverse-longitudinal momentum-space anisotropy of the plasma can be large. In order to accomplish this, one expands around an anisotropic background where the momentum-space anisotropies are built into the LO term
\begin{equation}
   f(x,p) = 
   f_{\rm aniso}\!\left(\frac{\sqrt{p^\mu \Xi_{\mu\nu}(x)p^\nu}}{\Lambda(x)}, \frac{\mu(x)}{\Lambda(x)}\right)
   + \delta\!\tilde{f}(x,p).
\label{eq:ahexp}
\end{equation}
Here $\Xi_{\mu\nu}$ is a second-rank tensor that measures the amount of momentum-space anisotropy and $\Lambda$ is the transverse temperature scale which can be identified with the true temperature of the system only in the isotropic equilibrium limit. $\mu(x)$ is the effective chemical potential of the particles. Specifically, LO anisotropic hydrodynamics ({\sc aHydro}) is based on an azimuthally symmetric ansatz for $\Xi_{\mu\nu}(x)$ \cite{Martinez:2010sc} involving a single anisotropy parameter $\xi$ such that $p^\mu \Xi_{\mu\nu}(x)p^\nu$ reduces to ${\bf p}^2 + \xi(x) p_L^2$ in the LRF. The leading-order LRF distribution thus becomes of Romatschke-Strickland (RS) form \cite{Romatschke:2003ms} which has spheroidal surfaces of constant occupation number. The dynamical equations of {\sc aHydro} were derived from kinetic theory by taking $f(x,p) = f_{\rm aniso}(x,p)$ (i.e. by ignoring the correction $\delta\!\tilde{f}$ in Eq.~(\ref{eq:ahexp})), and using the zeroth and first moments of the Boltzmann equation in the relaxation time approximation \cite{Martinez:2010sc,Martinez:2012tu}.  

Recently, the corrections due to $\delta\!\tilde{f}$ were included in a next-leading-order treatment of anisotropic hydrodynamics dubbed ``viscous anisotropic hydrodynamics'' ({\sc vaHydro}) \cite{Bazow:2013ifa}.  In the {\sc vaHydro} framework, dissipative effects due to the spheroidally deformed $f_{\rm aniso}$ are treated non-perturbatively, while the corrections $\delta\!\tilde{f}$ are treated perturbatively.  Another interesting recent development has been to generalize the RS form from spheroidal to ellipsoidal form at leading order \cite{Tinti:2013vba}, at least for the case of a system which possesses cylindrical symmetry in space.  This development offers some promise to treat all diagonal components of the energy-momentum tensor non-perturbatively, while treating only the off-diagonal components perturbatively.  It will be very interesting to see how this plays out.

At leading order in the formalism, the {\sc aHydro} dynamical equations result from taking moments of the Boltzmann distribution.  It was shown that for (0+1)d systems the resulting dynamical equations reduce to the Israel-Stewart equations of second-order viscous hydrodynamics in the limit of small momentum-space anisotropies \cite{Martinez:2010sc}.  In addition, and perhaps more importantly, the equations are able to reproduce the large shear viscosity to entropy density ratio limit ($\bar\eta \equiv \eta/{\cal S} \rightarrow \infty$) which gives longitudinal free streaming in the case of (0+1)d dynamics \cite{Martinez:2010sc}. The leading-order formalism has been extended to describe the full 3+1d dynamics of an spheroidally anisotropic QGP in the conformal (massless) limit \cite{Ryblewski:2011aq,Florkowski:2011jg,Martinez:2012tu,Ryblewski:2012rr}.  It has also been extend to describe two-component systems consisting of quarks and gluons with explicit baryon number conservation~\cite{Florkowski:2012as,Florkowski:2013uqa}.

At next-to-leading order, one can include deviations from the spheroidal leader-order form that arise from $\delta\!\tilde{f}$ in Eq.~\ref{eq:ahexp}.  The resulting framework provides non-perturbative dynamical equations for the momentum-space anisotropy ($\xi$), the transverse temperature ($\Lambda$), the LRF velocity (${\bf u}$), and the longitudinal boost-angle associated with the LRF ($\vartheta$).  These equations are coupled to an evolution equation for the dissipative corrections generated by $\delta\!\tilde{f}$.  The resulting dynamical equations for a conformal system in the relaxation time approximation are 
\begin{eqnarray}
\label{eq:de1}
&& \hspace{-7.5cm} {\rm Zeroth\;moment:} \nonumber \\
&& \nonumber \\
\frac{\dot{\xi}}{1+\xi}-6\frac{\dot{\Lambda}}{\Lambda}-2\theta &=&
2\Gamma\left(1-\sqrt{1{+}\xi}\,{\cal R}^{3/4}(\xi)\right) ,
\end{eqnarray}
\vspace{-5mm}
\begin{subequations}
\label{eq:de2}
\begin{eqnarray}
\label{eq:de2a}
&& \hspace{-1.1cm} {\rm First\;moment:} \nonumber \\
&& \nonumber \\
&&{\cal R}' \dot\xi + 4 {\cal R}\frac{\dot\Lambda}{\Lambda} = 
- \left({\cal R}{+}\frac{1}{3} {\cal R}_\perp\right) \theta_\perp
- \left({\cal R}{+}\frac{1}{3} {\cal R}_L\right) \frac{u_0}{\tau} 
+\frac{\pit^{\mu\nu}\sigma_{\mu\nu}}{{\cal E}_0(\Lambda)} ,
\\
\label{eq:de2b}
&&\left[3{\cal R}{+}{\cal R}_\perp\right]\dot{u}_\perp = 
-{\cal R}_\perp' \partial_\perp \xi 
- 4  {\cal R}_\perp \frac{\partial_\perp\Lambda}{\Lambda}
-u_\perp\Bigl({\cal R}_\perp' \dot\xi{+}4{\cal R}_\perp \frac{\dot\Lambda}{\Lambda}\Bigr)
\nonumber\\
&&\hspace*{2.8cm}
-u_\perp({\cal R}_\perp{-}{\cal R}_L)\frac{u_0}{\tau}  
+ \frac{3}{{\cal E}_0(\Lambda)}
\left(\frac{u_x\Delta^1_{\ \nu}+u_y\Delta^2_{\ \nu}}{u_\perp}\right) \partial_\mu\pit^{\mu\nu} ,
\\
\label{eq:de2c}
&&\left[3{\cal R}{+}{\cal R}_\perp\right] u_\perp\dot{\phi}_u = 
-{\cal R}_\perp' D_\perp\xi 
- 4 {\cal R}_\perp \frac{D_\perp\Lambda}{\Lambda}
-\frac{3}{{\cal E}_0(\Lambda)}\left(\frac{u_y\partial_\mu\pit^{\mu 1}-u_x\partial_\mu\pit^{\mu 2}}{u_\perp}\right) , \qquad
\end{eqnarray}
\end{subequations}
\vspace{-5mm}
\begin{eqnarray}
&& \hspace{-1.6cm} {\rm Second\;moment:} \nonumber \\
&& \nonumber \\
\dot{\pit}^{\mu \nu} &=& - 2\dot{u}_\alpha \pit^{\alpha(\mu} u^{\nu)}
-\Gamma\Bigl[
\bigl({\cal P}(\Lambda,\xi){-}{\cal P}_\perp(\Lambda,\xi)\bigr)\Delta^{\mu\nu}
+\bigl({\cal P}_{\rm L}(\Lambda,\xi){-}{\cal P}_\perp(\Lambda,\xi)\bigr) z^\mu z^\nu
+\pit^{\mu\nu}
\Bigr]
\nonumber \\
&& \hspace{0.5cm}
+{\cal K}^{\mu\nu}_0+{\cal L}^{\mu\nu}_0 + \, {\cal H}^{\mu\nu\lambda}_0\dot{z}_\lambda
+{\cal Q}^{\mu\nu\lambda\alpha}_0\nabla_\lambda u_\alpha 
+{\cal X}^{\mu\nu\lambda}_0 u^\alpha\nabla_\lambda z_\alpha
\nonumber \\
&& \hspace{1cm}
-2\lambda^0_{\pi\pi}\pit^{\lambda\langle\mu}\sigma_{\ \lambda }^{\nu\rangle}
+2\pit^{\lambda\langle\mu}\omega_{\ \lambda}^{\nu\rangle} 
-2\delta^0_{\pi\pi}\pit^{\mu\nu} \theta \, .
\label{eq:pimunu}
\end{eqnarray}
Above primes indicate a derivative with respect to $\xi$, 
dots indicate a comoving derivative, i.e. $\dot{f} = D f = u^\mu \partial_\mu f$,
$D_\perp = {\bf u}_\perp \cdot {\bf \nabla}_\perp$ is the transverse comoving derivative,
$\theta = \partial_\mu u^\mu$ is the expansion scalar, $\theta_\perp = {\bf \nabla}_\perp\cdot {\bf u}_\perp$ 
is the transverse expansion scalar, and $\phi_u=\tan^{-1}(u_y/u_x)$.  The functions ${\cal R}$, 
${\cal R}_\perp$, and ${\cal R}_L$ are analytically known nonlinear functions that arise in 
the calculation of the anisotropic thermodynamical functions ${\cal E}$, ${\cal P}_T$, and ${\cal P}_L$ \cite{Martinez:2010sc}.  The isotropic 
thermodynamical functions ${\cal E}_0$ and ${\cal P}_0$ can be calculated once the form of the underlying 
isotropic distribution function which enters the RS form is specified.  The parentheses and square brackets
indicate symmetrization and antisymmetrization, respectively.  Angular brackets indicate projection of the transverse and traceless tensor components.  
Above $\Gamma = {\cal R}^{1/4}(\xi)\Lambda/(5\bar{\eta})$ is the local relaxation rate which
depends on the degree of anisotropy, the local transverse temperature, and the shear viscosity to entropy ratio.  The dissipative forces ${\cal K}_0$, ${\cal L}^{\mu\nu}_0$ etc. and transport coefficients $\lambda^0_{\pi\pi}$ and $\delta^0_{\pi\pi}$ appearing in Eq.~(\ref{eq:pimunu}) are tabulated in Ref.~\cite{Bazow:2013ifa}.
Finally, $\Delta^{\mu\nu}=g^{\mu\nu}{-}u^\mu u^\nu$ is a transverse projector, $\sigma^{\mu\nu}\equiv\nabla^{\langle\mu}u^{\nu\rangle}$ is velocity shear tensor, $\omega^{\mu\nu}\equiv\nabla^{[\mu}u^{\nu]}$ is the vorticity tensor, and $z^\mu$ is the anisotropy direction, which in the LRF is given by $z^\mu = (0,0,0,1)$.
Note that, if one takes the non-spheroidal shear $\pit^{\mu\nu}$ to zero, then the zeroth and first moments reduce to the leading-order anisotropic hydrodynamics equations.

\begin{figure}[t]
\begin{center}
\includegraphics[width=0.49\textwidth]{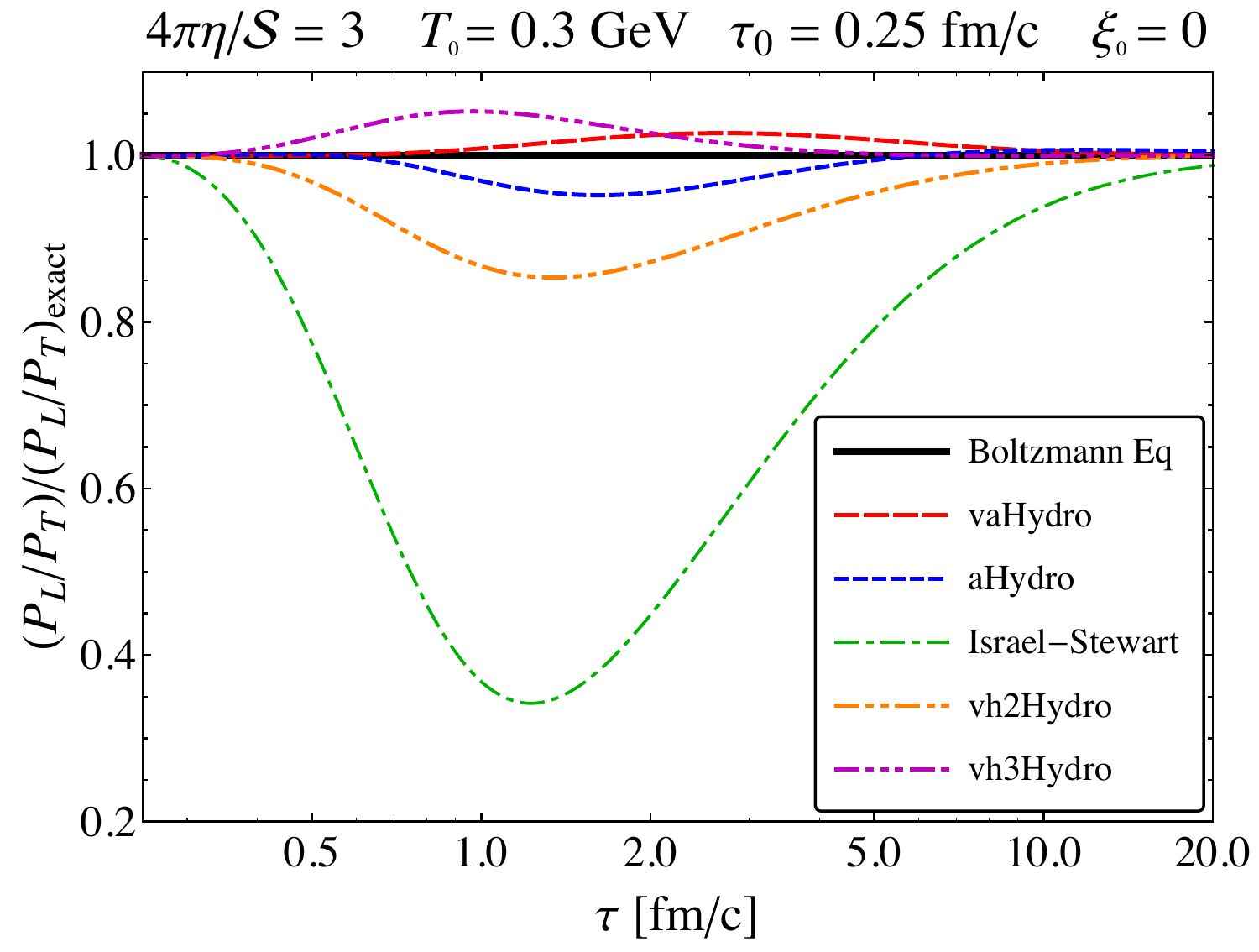}
\includegraphics[width=0.49\textwidth]{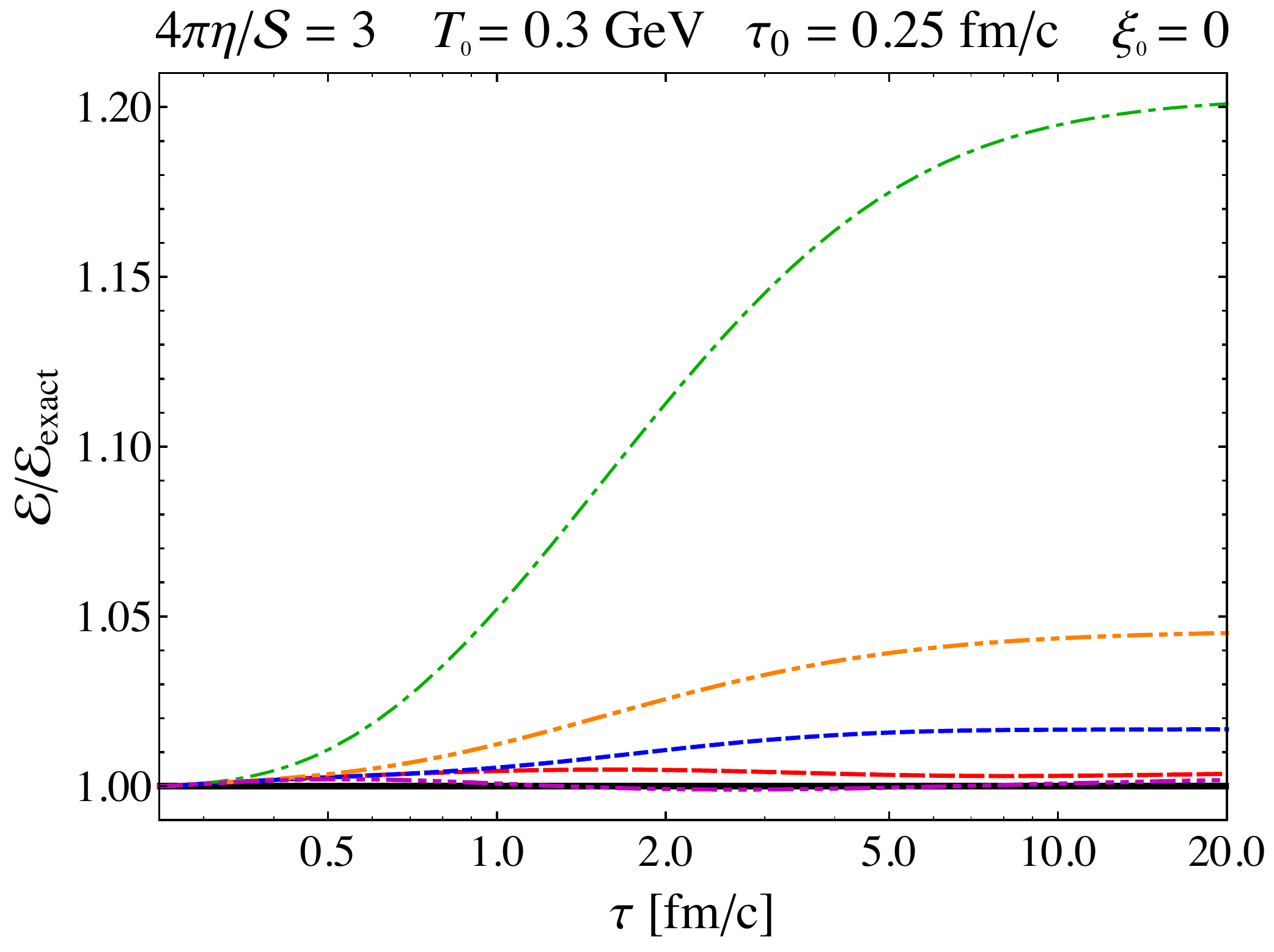}
\end{center}
\vspace{-7mm}
\caption{(Color online) Relative error of various dissipative hydrodynamics approaches compared to the exact solution of the (0+1)d Boltzmann equation in relaxation time approximation.  The left panel shows the relative error in the pressure ratio and the right panel shows the relative error in the energy density.  The legend applies to both panels and the various approximations are described in the text.}
\label{fig:rtacomp}
\end{figure}

As a way to test to the efficacy of this approach, one can check how well the equations above reproduce exact 
numerical solution of the Boltzmann equation in the case of transversely homogeneous boost invariant (0+1)d 
system \cite{Florkowski:2013lza,Florkowski:2013lya}.  In Fig.~\ref{fig:rtacomp} I plot the longitudinal to 
transverse pressure ratio scaled by the exact solution (left) and the energy density scaled by exact solution 
(right).  For all approximations shown the system was initialized isotropically $\xi_0=0$ with an initial
temperature of $T_0 = 300$ MeV at $\tau_0$ = 0.25 fm/c and I assumed $4\pi\bar\eta = 3$.  The solid black line
is the scaled exact solution for reference, the long-dashed red line is the {\sc vaHydro} result, the short-dashed
blue line is the {\sc aHydro} result, the green dot-dashed line is the Israel-Stewart second-order viscous 
hydrodynamics result, the orange dot-dot-dashed line is the complete second order viscous hydrodynamics 
result of Denicol et al. \cite{Denicol:2010xn,Denicol:2012cn,Denicol:2012es}, and the purple dot-dot-dot-dashed
line is the third order Chapman-Enskog viscous hydrodynamics result of Jaiswal \cite{Jaiswal:2013vta}.  As
can be seen from Fig.~\ref{fig:rtacomp} the recently obtained second-order anisotropic dynamics equations
({\sc vaHydro}) are superior to all other second-order approaches and comparable the third-order treatment of Jaiswal. 
In Ref.~\cite{Bazow:2013ifa} a more comprehensive study of the initial condition dependence of these
quantities and particle production demonstrated that {\sc vaHydro} best reproduces the exact 
solution to the Boltzmann equation.  We note, importantly, that out of all of the approximations considered,
the Israel-Stewart approximation is the poorest approximation considered.

%%%%%%%%%%%%%%%%%%%%%%%%%%%%%%%%%%%%%%%%%%%%%%%%%%%%%%%%%%%%%%%%%%%%%%%%%%%%%%%%%%%%%%%%%%%%%%%%%%%%%%%%%%%%%%
\section{Conclusions and Outlook}
%%%%%%%%%%%%%%%%%%%%%%%%%%%%%%%%%%%%%%%%%%%%%%%%%%%%%%%%%%%%%%%%%%%%%%%%%%%%%%%%%%%%%%%%%%%%%%%%%%%%%%%%%%%%%%

Dissipative hydrodynamical models are able to describe the collective flow of the QGP produced at RHIC and LHC, both in terms of event-averaged observables and their underlying probability distributions, with a surprising level of accuracy.  Since dissipative hydrodynamics implies the existence of momentum-space anisotropies in the QGP, one must now conclude, based on empirical evidence alone, that the QGP might be thermal but strongly anisotropic in momentum-space, implying that the QGP has (at least) two temperatures, a transverse one and a longitudinal one.  This means, in practice, that one has to fold the momentum-space anisotropy of the underlying one-particle parton distribution functions into the calculation of various processes.  There have been some initial work along these lines (see Ref.~\cite{Strickland:2013uga} for a collection of relevant references), but there is much work left to do.  

On the dynamics front, the recently developed {\sc vaHydro} approach provides a complete second-order treatment which takes into account plasma anisotropies from the outset and, as a result, yields a superior approximation scheme.  Future developments will include implementation of {\sc vaHydro} numerical codes.  Since the expansion around a locally anisotropic momentum distribution results in smaller deviations $\delta\!\tilde{f}$ of the distribution function from the leading-order ansatz, the {\sc vaHydro} framework should yield results that are quantitatively more reliable, particularly when it comes to the early stages of QGP hydrodynamical evolution and near the transverse edges of the overlap region where the system is approximately free streaming.  Finally, as mentioned previously, another important recent development has been the development of leading-order ellipsoidal anisotropic hydrodynamics \cite{Tinti:2013vba}.  If the off-diagonal contributions to the energy-momentum tensor can be taken into account using similar methods as {\sc vaHydro}, this should yield an even better approximation scheme.

%%%%%%%%%%%%%%%%%%%%%%%%%%%%%%%%%%%%%%%%%%%%%%%%%%%%%%%%%%%%%%%%%%%%%%%%%%%%%%%%%%%%%%%%%%%%%%%%%%%%%%%%%%%%%%
\section*{Acknowledgments}
%%%%%%%%%%%%%%%%%%%%%%%%%%%%%%%%%%%%%%%%%%%%%%%%%%%%%%%%%%%%%%%%%%%%%%%%%%%%%%%%%%%%%%%%%%%%%%%%%%%%%%%%%%%%%%

\vspace{-2mm}
I thank my collaborators D. Bazow, U. Heinz, R. Maj, M. Martinez, R. Ryblewski, and W. Florkowski.
This work was supported in part by DOE Grant No.~DE-SC0004104.

%%%%%%%%%%%%%%%%%%%%%%%%%%%%%%%%%%%%%%%%%%%%%%%%%%%%%%%%%%%%%%%%%%%%%%%%%%%%%%%%%%%%%%%%%%%%%%%%%%%%%%%%%%%%%%
%\bibliographystyle{elsart-num}
\bibliographystyle{utphys}
\bibliography{strickland}
%%%%%%%%%%%%%%%%%%%%%%%%%%%%%%%%%%%%%%%%%%%%%%%%%%%%%%%%%%%%%%%%%%%%%%%%%%%%%%%%%%%%%%%%%%%%%%%%%%%%%%%%%%%%%%

\end{document}